\documentclass[aps,twocolumn,prl,graphicx]{revtex4}
\usepackage{graphicx}% Include figure files
\usepackage{dcolumn}% Align table columns on decimal point
\usepackage{bm}% bold math
\begin{document}

\title{Tailoring 10-nm-scale Suspended Graphene Junctions and Quantum Dots}
\author{V. Tayari, A.~C. McRae, S. Yi\u{g}en, J.~O. Island, J.~M. Porter, and A.~R. Champagne}
\email[]{a.champagne@concordia.ca}
\affiliation{Department of Physics, Concordia University, Montr\'{e}al, Qu\'{e}bec, H4B 1R6, Canada}

\begin{abstract}
The possibility to make 10-nm-scale, and low-disorder, suspended graphene devices would open up many possibilities to study and make use of strongly coupled quantum electronics, quantum mechanics and optics. We present a versatile method, based on the electromigration of gold-on-graphene bow-tie bridges, to fabricate low-disorder suspended graphene junctions and quantum dots with lengths ranging from 6 nm up to 55 nm. We control the length of the junctions, and shape of their gold contacts by adjusting the power at which the electromigration process is allowed to avalanche. Using carefully engineered gold contacts and a non-uniform downward electrostatic force, we can controllably tear the width of suspended graphene channels from over 100 nm down to 27 nm. We demonstrate that this lateral confinement creates high-quality suspended quantum dots. This fabrication method could be extended to other two-dimensional materials.
\end{abstract}

\maketitle %\maketitle must follow title, authors, abstract and \pacs

Graphene has very few intrinsic defects \cite{Meyer08} and, when suspended, its extrinsic bulk disorder can be annealed by Joule heating \cite{Bolotin08, Du08, Yigen13}. In nanoscale graphene devices, both the electronic \cite{Ponomarenko08,Young09, Wu12, Guttinger12, Engels13, Barreiro12} and mechanical \cite{vonOppen09, Weber14} degrees of freedom can be quantized. Ultra-small graphene devices have been fabricated on substrate \cite{Shi11, Barreiro12, Guttinger12}, but substrate disorder and fabrication residues affect their electronics and optics, while the clamping of the substrate restricts their mechanics. Ultra-small and low-disorder suspended graphene devices (SGDs) would be a very powerful platform to explore strongly interacting quantum electro-mechanics \cite{Fogler08,vonOppen09,Rocheleau10, Weber14} and electro-optics \cite{Freitag13, Britnell13, Kim14}. For example, ten-nanometer-scale SGDs would enable strain-engineering (up to $>$ 10$\%$ strain) of ballistic Dirac fermion transport \cite{Fogler08,Guinea10,Low10} using a mechanical breakjunction method \cite{Champagne05, Parks10}, allow the study of valleytronics \cite{Low10} and development of strain-transistors \cite{Fogler08}. Both the maximum operating frequency of graphene transistors \cite{Wu12,Zheng13} and the responsivity of graphene photodetectors \cite{Freitag13} increase in shorter devices with low disorder. Thus, ultra-small SGDs would be an ideal testing platform to develop THz frequency graphene electronics \cite{Zheng13}, sensitive bolometers \cite{Fong13,Kim14} and optical modulators\cite{Kim11}. To the best of our knowledge, the shortest low-disorder on-substrate graphene channels studied so far were approximately 40 nm long \cite{Wu12_2}. On substrate 10-nm-scale graphene devices have been fabricated \cite{Moser09,Lu10,Shi11, Barreiro12}, but these devices were highly disordered. Currently, there is no established fabrication method to create 10-nm-scale SGDs with low disorder. Here, we report such a procedure.

\begin{figure}
  \includegraphics[width=3.25in]{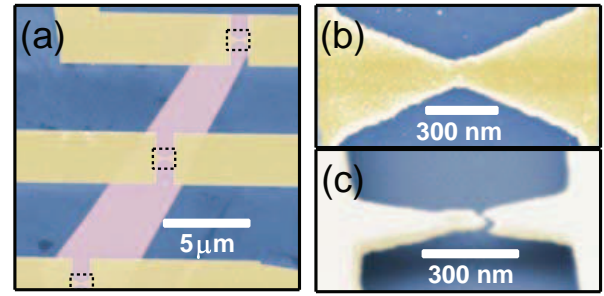}
  \caption{\label{}Fabrication of suspended gold-on-graphene breakjunctions. (a) Colorized SEM image showing gold contacts (yellow) connecting breakjunctions (dashed boxes) defined on a graphene crystal (purple). (b) Enlarged view of one bowtie-shaped breakjunction. The junctions are typically between 100 and 150 nm wide at their center. A reactive ion etch was used to remove the graphene crystal everywhere except under the gold breakjunction. (c) Tilted-SEM image of a breakjunction after a wet SiO$_2$ etch used to suspend the gold bridge, and a controlled gold electromigration which uncovered a nm-sized graphene channel. The graphene channel is not visible at this tilt angle, but similar channels can be seen in Figure 2.}
\end{figure}

\begin{figure}
\includegraphics[width=3.25in]{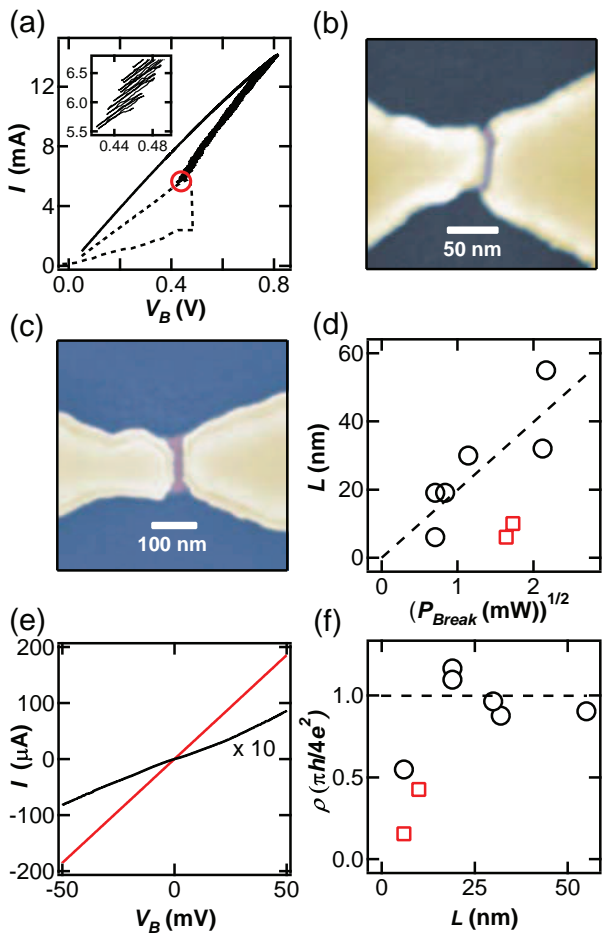}
\caption{\label{}Low-power EM, length control of the suspended junctions. (a) $I-V_{B}$ characteristics of the EM process for Device A, whose SEM image is shown in (b). The EM is done in two stages. In stage one (solid line), a feedback-controlled $V_{B}$ is applied to gradually etch the junction (see text for details). This process can be stopped at any position on the $I-V_{B}$ curve (red circle). In stage two, the junction is broken with a continuous $V_{B}$ ramp (dashed data). The breaking power, $P_{Break}$ (power at the red circle), controls the size and shape of the junction. When $P_{Break} <$ 5 mW, the junctions are roughly symmetric. The breaking power for Device A was 2.7 mW, and the created channel was $\approx$ 6 nm long as shown in (b). (c) SEM image of Device B, whose breaking power was 4.5 mW and channel length $\approx$ 26 nm. (d) Length, $L$, versus $P_{Break}^{1/2}$ for $P_{Break} <$ 5 mW devices. The black circles and red squares data are for devices electromigrated at $T =$ 4.2 and 1.5 K respectively. The data show a roughly linear dependence on $P_{Break}^{1/2}$ for $L$ from 6 nm to 55 nm. (e) $I-V_{B}$ transport data at 4.2 K in Devices A (red data) and B (black data). The curves are nearly linear as expected for low-disorder channels. (f) The dc resistivity, $\rho$, of the devices in (d) versus $L$. We expect the very short ($L <$ 15 nm) channels to be significantly doped by charge transfer from the contacts \cite{Giovannetti08, Golizadeh09}. In the longer channels, $\rho$ is close to the expected value for undoped ballistic graphene\cite{Tworzydlo06}, $\rho_{o} = \pi h/4e^{2}$.}
\end{figure}

We show that using electromigration (EM), we can tailor the shape of suspended gold breakjunctions deposited on graphene without damaging the graphene. This allows us to make low-disorder SGDs whose channel lengths range from $\approx$ 6 to 55 nm, and resistivities are close to the ballistic limit. For a fixed breakjunction geometry, the power $P_{Break}$ at which the EM is allowed to avalanche (i.e. no external feedback) controls the length of the SGDs. We also demonstrate that using a relatively elevated $P_{Break}$ and asymmetric dc biasing (electric field), the EM produces highly asymmetric source and drain electrodes. We prepare needle-shaped source electrodes, and then apply a non-uniform downward electrostatic force to the suspended channels to tear them into very narrow channels ($\leq$ 30 nm wide). The lateral confinement in such narrow channels creates a band gap \cite{Han07, Guttinger12}. We observe clear Coulomb blockaded electron transport at low temperature in these devices, demonstrating that they form suspended quantum dots (QDs). The QD area we extract from transport data matches the area of the suspended channel measured by SEM imaging, confirming that the EM procedure does not introduce bulk defects in the SGD. Electrostatic tearing of suspended graphene has been predicted to lead to atomically-ordered graphene edges \cite{Moura13}. Edge-ordered suspended graphene QDs would be of great interest as electro-mechanical qubits \cite{Rocheleau10, Engels13}. Our SGD fabrication procedure is able to tailor both the length and width of 10-nm scale low-disorder suspended graphene devices. It offers the potential to explore a wide range of graphene physics in both ballistic and tunneling transport devices, and should be extendable to other 2-dimensional materials (e.g. MoS$_2$, WSe$_2$).

Figure 1 shows colorized SEM images of the different fabrication stages we followed to make suspended gold-on-graphene breakjunctions. We start with heavily-doped Si wafers with a 300 nm-thick SiO$_{2}$ film on their top side. The Si substrate is used as a gate electrode. Figure 1a shows in purple a large graphene crystal exfoliated on the SiO$_2$/Si substrate. We confirmed that our exfoliated crystals are single-layer by optical and Raman microscopy. On top of the graphene flakes, using e-beam lithography, we defined bow-tie shaped gold breakjunctions (40 nm thick Au, no adhesion layer) connected by large gold wires (3 $\mu$m wide). The dashed boxes in Figure 1a highlight the breakjunctions, one of which is enlarged in Figure 1b. The junctions are 1 $\mu$m long and are typically 100 to 150 nm wide at their narrowest point. After fabricating the breakjunctions, we used an O$_2$ reactive ion etch to remove the graphene crystal everywhere except under the gold junctions. Next, we suspended the gold-on-graphene junctions by etching away the SiO$_2$ ($\approx$ 50 nm deep) underneath the junctions using a wet buffered oxide etch. The depth of the etch is adjusted to ensure that the breakjunctions are fully suspended. Figure 1c shows a suspended junction. The small gap at the center of the gold bridge was made during the final step of the fabrication. We used a feedback-controlled EM (described below) to open a nm-sized gap in the gold junction without damaging the graphene film underneath. The graphene channel is not visible in tilted-SEM imaging, but similar channels are visible in Figure 2.

Figure 2 describes the low power gold breakjunction EM procedure we used to define 6 nm to 55 nm long graphene channels with nearly symmetric contacts. This procedure builds on previous work on breakjunction electromigration \cite{Park99, Champagne05, Esen05}. We previously described a related prodecure to create defect free ultra-short single-wall carbon nanotube transitors and QDs \cite{Island11}, and NEMS \cite{Island12}. One key ingredient of our method is that gold EM takes place at a temperature of a few hundred degrees Celsius \cite{Jeong14} (and the absence of a substrate limits the formation of hot spots\cite{Jeong14}). Coincidentally, this gold-EM temperature is perfectly suited to anneal adsorbed impurities on suspended graphene devices \cite{Bolotin08} under high (cryogenic) vacuum. Yet, the EM temperature is much too low to create defects in graphene \cite{Dorgan13}.

Figure 2a shows the current versus bias voltage ($I-V_{B}$) trace during the gold EM of Device A. The resulting SGD is shown in Figure 2b. We used dc voltage biasing, with $V_{B}$ applied to the source side of the junction while the drain side is grounded. The EM is typically done in two stages. In stage one (solid data in Figure 2a), we gradually etch the gold junction to narrow it down. $V_{B}$ is ramped up and the device's resistance is monitored by a feedback control \cite{Esen05}. When the junction's resistance increases (narrowing of the gold constriction) the feedback partially ramps down $V_{B}$ to reduce the power dissipated in the junction and thus its etching rate \cite{Esen05}. This process of slightly ramping up and down $V_{B}$, to gradually increase the resistance of the junction, is repeated many times until the junction reaches some desired $I-V_{B}$ position (red circle in Figure 2a) which corresponds to a power $P_{Break} = I \times V_{B}$. The inset of Figure 2a shows a zoom-in on the solid black trace, where the up and down voltage ramps are visible. Once the desired $I-V_{B}$ position is reached, $V_{B}$ is ramped back down to zero. During the second stage of the EM (dashed data in Figure 2a) we apply a single continuous $V_{B}$ ramp until the junction breaks, after which we rapidly ramp down the voltage. This final break happens very close to the last $I-V_{B}$ position of stage one (red circle in Figure 2a), and thus permits us to break the junction with a prescribed $P_{Break}$. For example, the junctions of Devices A and B in Figure 2b,c broke at $P_{Break} =$ 2.7 mW and 4.5 mW respectively, and have channel lengths, $L =$ 6 $\pm$ 3 nm and 26 $\pm$ 3 nm.

All the junctions reported in Figure 2 were fabricated with $P_{Break} <$ 5 mW, which yields nearly symmetric source and drain electrodes, and channels with a well-defined length. Such symmetric contacts mean that we can ascribe a well defined injection angle into the channel to the charge carriers. This would permit quantitative studies of ballistic electron transport in ultra-short SGDs\cite{Fogler08}. The rate of gold EM is both dependent on temperature\cite{Esen05} and the momentum transferred to gold atoms from the current flow\cite{Park99}. To predict the dependence of $L$ of the graphene junctions on $P_{Break}$ would be complex as it depends on the gold EM rate, but also on the local thermal conductivity of gold and graphene, and the diffusion rate of gold on graphene. We observe empirically that $L$ of the graphene channels we produce scales like $P_{Break}^{1/2}$ at a constant cryostat $T$. Figure 2d shows $L$, measured by SEM, of the SGDs fabricated at low power as a function of $P_{Break}^{1/2}$. The data in black circles and red squares are respectively for junctions which were electromigrated at $T =$ 4.2 and 1.5 K. We note that for a similar $P_{Break}$, the junctions broken at lower $T$ are shorter. The scatter in the data is most likely explained by microscopic disorder in the gold film. Figure 2e shows $I-V_{B}$ transport data (all transport data reported in this study were taken at $T =$ 4.2 K) in Devices A (red) and B (black). We note that the $I-V_{B}$ trace for Device A is precisely linear, and the very small non-linearity for Device B may be due to lateral confinement. Both the shapes of these $I-V_{B}$ traces and the value of their slopes ($R^{-1} = dI/dV_{B}$) strongly suggest that the graphene channels do not have significant bulk disorder or doping. Figure 2f displays the resistivity $\rho = R\times W/L$ vs.\ $L$ for the samples in Figure 2d. The measured $\rho$ is close to the expected ballistic resistivity for undoped graphene \cite{Tworzydlo06,DasSarma11}, $\rho_{o} = \pi h/4e^{2}$. This confirms that the EM anneals the adsorbed impurities on the suspended graphene very effectively without damaging the crystal. The resistivities of the shortest devices ($L <$ 15 nm) are smaller than $\rho_{o}$ as expected, since in ultra-short channels the extrinsic charge doping from the gold contacts \cite{Giovannetti08, Golizadeh09} affects the entire length of the channel. We note that the universal value of $\rho_{o}$ is only valid for truly ballistic transport which cannot be described by a diffusive Boltzmann approach \cite{Tan07, DasSarma11}, and in the limit of devices with a large aspect ratio\cite{Tworzydlo06} $W/L\geq$ 4 which roughly applies to all of the devices in Figure 2f. The symmetric, low-disorder, and ultra-short SGDs presented in Figure 2 are of interest to develop extremely high frequency and sensitive NEMS\cite{Bunch07,Island12,Chen13} and explore the ballistic transport of Dirac fermions in graphene\cite{Fogler08, Low10}.

\begin{figure}
    \includegraphics [width=3.25in]{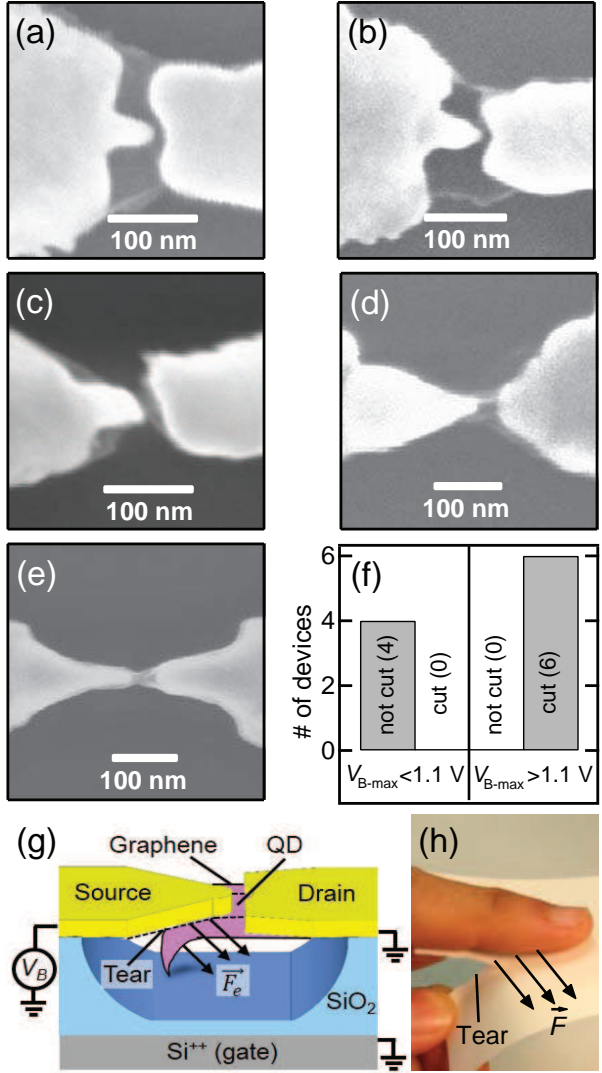}
    \caption{\label{} High-power EM, asymmetric contacts and controlled tearing of suspended graphene. (a)-(e) SEM images of Devices C, D, E, F, and G respectively. These junctions were electromigrated with high powers ($P_{Break} =$ 14, 12.1, 13.2, 18, and 14.4 mW), creating highly asymmetric gold contacts. In Figure 3a,b the suspended graphene channels (light gray) were left untouched by the gold EM. On the other hand, in (c)-(e) the graphene channels were cut down to the width of their sharp source contacts. A $V_{B} \geq 1.1$ V was applied (during or after EM) to the channels which were torn. (f) Histogram of the devices with $L \geq$ 15 nm, showing which samples had their channel cut or not. The devices to which the applied $V_{B}$ exceeded 1.1 V had their channels cut, while we did not observe any tear in devices whose $V_{B}$ was kept below 1.1 V. (g) Diagram showing the proposed graphene tearing mechanism. We apply a non-uniform downward electrostatic force to the channel near the source electrode. A tear can nucleate from a defect at the edge of the flake, and be guided along the length of the channel by the clamping of the needle-shaped source electrode. (h) The method we use to tear graphene is very similar to the usual approach used to tear a sheet of paper.}
\end{figure}

Figure 3 shows that we can fabricate junctions with highly-asymmetric source and drain electrodes using a high-power EM ($P_{Break} >$ 10 mW). We also demonstrate that using narrow source electrodes and a localized downward electrostatic force at the edge of these contacts, we can tear the SGDs into narrower channels. A higher EM power means a higher junction temperature and a much faster EM rate \cite{Esen05}. Combined with dc biasing, which produces a unidirectional electron momentum transfer to the gold atoms, it allows us to control the gold electrodes' asymmetry. When $P_{Break}$ exceeds 10 mW, the created junctions are highly asymmetric as can be seen in Figure 3a-e for Devices C, D, E, F, and G ($P_{Break} = $ 14, 12.1, 13.2, 18, and 14.4 mW). The source electrodes in Figure 3 take a needle shape while the drain electrodes remains wider. For comparison, Figure 2b,c show much more symmetric electrodes for Devices A and B whose $P_{Break}$ were 2.7 and 4.5 mW. Suspended graphene is visible in Figure 3a-e as the light grey contrast. We show raw SEM images in order to reveal the details of the shape and topography of the graphene channels. We observe that the graphene channels were unperturbed by the gold EM in Figure 3a,b and span the entire width of the original gold bow-tie junctions. On the other hand, in Figure 3c-e the suspended channels were torn from their original width down to the width of their needle-shaped source electrodes. The channels were narrowed down by as much as a factor of 4 from the original breakjunctions' width.

We did not observe channel tearing in SGDs with $L <$ 15 nm. Such short channels require a low power EM and have rather symmetric electrodes (see Figure 2), indicating that the asymmetric shape of the electrodes plays an important role in the tearing process. Figure 3f summarizes the occurrence of channel cutting in devices with $L >$ 15 nm. In samples whose applied $V_{B}$ exceeded 1.1 V, either during the last stage of the EM (dashed data in Figure 2a) or after the EM (additional $V_{B}$ ramp), we observe that the channels are cut down to the width of the needle-shaped source electrodes (Figure 3c-e). The maximum bias, $V_{B-max}$, applied to Devices E, F, and G in Figure 3c-e were respectively 1.12, 1.22 and 2.0 V. When $V_{B}$ was kept below 1.1 V at all steps of the preparation and measurement of the samples, we did not observe channel tearing in SEM imaging (Figure 3a,b). The specific value of the $V_{B-max}$ threshold is an empirical observation, and is most likely dependent on the geometry of our breakjunctions. This graphene tearing capability allows us to create very narrow suspended devices, for instance Devices E, F, and G are respectively 44, 38 and 27 nm wide.

Figure 3g shows the mechanism we propose to explain the tearing of the suspended channels. It has been shown that a downward electrostatic force applied to suspended graphene can propagate a tear from a defect on the edge of the crystal\cite{Moura13}. Since our suspended graphene channels were etched using reactive ion etching, their edges have a high density of defects. We showed in Figure 2 that the suspended channels have nearly zero intrinsic doping, and thus the charge densities in the channels are directly dependent on the electrostatic potential. The biasing condition of our devices is such that $V_{B}$ is applied to the source electrode while both the drain and gate are grounded. This special configuration creates a non-uniform electric field and electrostatic force localized at the graphene near the source electrode (heavily doped by $V_{B} > 1.1$ V) pulling it toward both the gate and drain electrode (see Figure 3g). The force on the suspended graphene near the drain (ground) electrode is very small as its charge density is very small. The source electrode clamps down the graphene, and its needle shape can guide the propagation of a tear along the channel's length. Figure 3h shows that the geometry we use to tear our devices is very closely analogous to the approach usually taken to tear a sheet of paper.

We comment on why this electrostatic tearing was not previously observed in experiments on suspended graphene. The most likely tearing direction of suspended graphene is along a wrinkle or bend in the crystal \cite{Moura13}. Such a wrinkle can be created by the clamping of the contact electrodes. In previous experiments \cite{Bolotin08, Du08}, the contacts were rectangular and thus any tear would most likely have propagated perpendicularly to the channels and destroyed the suspended devices. The out-of-plane electrostatic force applied to the channels was uniform and due to a gate voltage $V_{G}$. In this configuration the force between the graphene and gate (substrate) is similar to the force between the plates of a parallel plate capacitor and scales as $V_{G}^{2}$. The $V_{G}$ applied to suspended graphene has been limited to a few volts, specifically to avoid damaging the samples, and the calculated force due to such $V_{G}$ was shown to be too small to propagate tears \cite{Moura13}. In our experiment, the out-of-plane electrostatic force is localized near the source electrode. This electrode has the appropriate geometry to guide a tear longitudinally along the channel. Finally, the magnitude of the force on the channel near the source is much larger than in a regular transport experiment (where $V_{B} \sim$ mV does not significantly dope the channel). We can calculate an upper and lower bound estimate for the induced charge density in the graphene near the source when $V_B =$ 1.1 V. An upper limit is to let the coupling between the source and channel to be ideal, and thus set the Fermi energy of the nearby graphene be 1.1 V. This corresponds to a density near the source of $n_{s} \sim$ 9 $\times$ 10$^{13}$ cm$^{-2}$. A gate voltage of $\sim$ 1800 V would be required to induce such a density. A lower limit for $n_{s}$ can be extracted from the transport data for Device G, which will be discussed below in Figure 4b. After the tearing process (which can only reduce the coupling between the channel and source), Device G forms a QD and its transport data show well defined Coulomb diamonds. The slope of the downward Coulomb thresholds (dashed lines in Figure 4b) give the ratio $-C_{G}/C_{S}$, i.e. the ratio of the channel-gate capacitance to the channel-source capacitance. We measure $C_{S} \approx$ 140 $C_{G}$, and thus a 1.1 V applied to the source would dope the channel (near the source) by as much as $V_G =$ 150 V. Using these two limits we can infer that $n_{s}$ in our devices is equivalent to at least the density induced by $V_{G} \sim$ few 100s of volts. The resulting out-of-plane force, $F_{e}\propto n_{s}$($V_{G}-V_{B}$), is equivalent to the force created by a $V_{G}$ of at least a few 10s of volts, which is an order of magnitude larger than in most previous experiments. This is consistent with calculations showing that the force created by a $V_{G}$ of a few 10s of volts could propagate a tear in suspended graphene \cite{Moura13}.

The tailoring of SGDs into very narrow channels allows us to control the lateral confinement of charge carriers and open band gaps in the suspended channels. This capability will be useful to engineer the energy dissipation in NEMS\cite{Weber14} and the sensitivity of graphene photodetectors\cite{Freitag13, Fong13}. Moreover, electrostatic tearing of graphene is predicted to lead to atomically-ordered edges \cite{Moura13}, which could open the door to detailed studies of edge-ordered suspended graphene QDs. Figure 4a shows the maximum resistivity, $\rho_{max}$ (extracted from $\rho$ - $V_{G}$ data as in Figure 4c), as a function of the width of the suspended channels, $W$, for the 6 samples we fabricated using electrostatic tearing. In graphene nanoribbons, a band gap forms between the valence and conduction band due to lateral confinement. The size of this band gap is inversely proportional to the width of the ribbon\cite{Han07, Guttinger12}, and in order for this lateral confinement to dominate transport, a device needs to have small aspect ratio $W/L$. At the interfaces between the torn (narrow) channel and the wider graphene under the gold, the bandgap gives rises to tunnel barriers \cite{Guttinger12}. For devices with $W >$ 50 nm, we find in Fig.\ 4a that $\rho_{max}$ is very close to $\rho_{o}$, as we found previously (Figure 2f) for junctions which were not electrostatically torn. This indicates that the tearing process does not add substantial disorder to the bulk of the devices. For devices with $W <$ 50 nm in Figure 4a, we note a clear increase of $\rho_{max}$ as $W$ decreases. The resistivities of our narrowest samples are lower than for lithographically defined QDs and ribbons of similar widths \cite{Han07, Guttinger12}, suggesting that they have a lower edge disorder.

Figure 4b shows $I-V_{B}$ transport data in Devices G (red data, $W$ = 27 nm) and H (black data, $W$ = 70 nm). The data are linear for Device H, showing no sign of charge confinement. For the much narrower Device G, the $I-V_{B}$ data are highly non-linear and show a clear Coulomb blockade. The strong connection between the width of our torn suspended ribbons and their transport characteristics is detailed in Figure 4c which shows the evolution of the conductivity, $\sigma$ vs. $V_{G}$ in Device G (red data, $W$ = 27 nm), Device F (blue data, $W$ = 38 nm), and Device H (black data, $W$ = 70 nm). The applied $V_{B}$ was 5 mV. While $\sigma$ of Device H is very weakly dependent on $V_{G}$ and approximately equal to $\sigma_{o} = 1/\rho_{o}$, both Devices F and G show much lower conductivity and large Coulomb oscillations, indicating that they behave as quantum dots. As expected, the overall conductivity of Device G (narrower) is lower, and shows larger oscillations, than for Device F (wider).

Figure 4d shows $dI/dV_{B}-V_B-V_G$ data for Device G at $T =$ 4.2 K. We observe the charge ground states of the QD as Coulomb diamonds (dashed lines) with a single set of positive and negative slopes as expected for a single QD. The diamonds in Figure 4d hint at a four-fold periodicity. The third diamonds are suppressed, which may indicate a Kondo resonance \cite{Goldhaber98}. So far, periodicity in the shell filling structure and the Kondo effect have not been observed in lithographically defined graphene QDs \cite{Guttinger12}, but were seen in ultra-narrow nanoribbons with low edge-disorder made from un-zipped nanotubes\cite{Wang11}. The energy needed to add an electron to the QD, $\Delta$, corresponds to the half-height of the Coulomb diamonds times $e$, the electron's charge. This addition energy has two components $\Delta = E_{C} + \delta$, where $E_{C}$ is the charging energy due to the total classical capacitance of the QD, and $\delta$ is the quantum energy level spacing. In graphene QDs smaller than about 50 nm in size, $\delta$ is expected to be of the same magnitude or larger than $E_{C}$ \cite{Yang07, Ponomarenko08, Guttinger12}, which is drastically different than in semiconducting QDs where $E_{C}$ dominates. A four-fold or two-fold degeneracy of energy levels is expected in edge-ordered graphene QDs due to the spin and valley degeneracies \cite{DasSarma11}. The taller diamonds in Figure 4d (labelled 1) are thus likely to correspond to the first electron added to an electronic shell, and the half-height of these diamonds is the sum of $E_C$ and $\delta$. The half-height of the other diamonds should correspond only to $E_C$ as additional electrons (labelled 2, 3, 4) are added to degenerate (or nearly degenerate) levels. We note that the four-fold periodicity in the addition energy is also visible in the $\sigma - V_{G}$ data in Figure 3c for both Devices F and G.

\begin{figure}
    \includegraphics [width=3.25in]{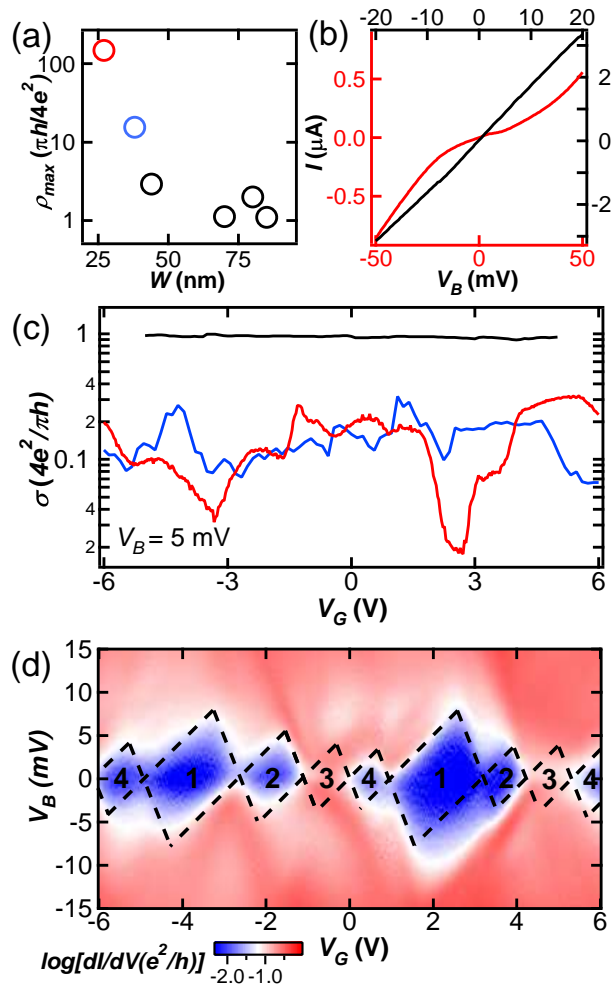}
    \caption{\label{} Electron transport in suspended QDs ($T =$ 4.2 K). (a) The maximum resistivity, $\rho_{max}$, (from (c) and similar data) vs. the width, $W$, of the suspended channels for the six devices we fabricated by electrostatic tearing. We observe a steady increase in $\rho_{max}$ when $W < $ 50 nm, which is consistent with low-disorder edges, and signals the onset of QD transport. (b) $I-V_{B}$ transport data in Devices G (red data, $W$ = 27 nm) and H (black data, $W$ = 70 nm), at $V_{G} = 0$. The data is linear for the Device H which shows no signs of charge confinement as expected. In much narrower devices, such as Device G, the $I-V_{B}$ data is highly non-linear (Coulomb blockade). (c) Conductivity, $\sigma$ vs.\ $V_{G}$ for Device G (red data, $W$ = 27 nm), Device F (blue data, $W$ = 38 nm), and Device H (black data, $W$ = 70 nm). The applied $V_{B}$ was 5 mV. While $\sigma$ of Device H is very weakly dependent on $V_{G}$, both Devices F and G show large Coulomb oscillations indicating that they behave as quantum dots. (d) Detailed charge transport $dI/dV_{B}-V_B-V_G$ data for Device G. From the half-height and full-width of its Coulomb diamonds (dashed lines) we extract a charging energy $E_{C}\approx$ 4 meV, energy level spacing $\delta \approx$ 4 meV, and gate capacitance $C_{G}\approx$ 1.1 $\times 10^{-19}$ F. The area of the QD extracted from $C_{G}$ closely matches the area of the suspended channel visible in Figure 3e, indicating that there is a single QD in the channel.}
\end{figure}

From Figure 4d, we extract $E_C \approx$ 4 meV and $\delta \approx$ 4 meV in Device G. The theoretical values of $\delta$ calculated from various theoretical model \cite{Yang07, Ponomarenko08,Guttinger12} for a QD of the size of the suspended channel in Device G (Figure 3e, $W =$ 27 nm and $L =$ 40 nm) are consistent with $\delta\sim$ 4 meV. From the full-width of the Coulomb diamonds in Figure 4d, we measure the value of $e/C_{G}$ where $C_{G} \approx$ 1.1 $\times 10^{-19}$ F is the QD-to-gate capacitance. Modeling $C_{G}$ as a parallel-plate capacitor, we extract from it a QD area, $A_{transport} =$ 1500 $\pm$ 350 nm$^{2}$ which matches well with the measured channel area in the SEM image of Figure 3e, $A_{SEM} =$ 1300 $\pm$ 100 nm$^2$, this is additional evidence that a single QD spans the entire suspended channel. This further confirms that our gold EM and electrostatic tearing procedures did not introduce bulk defects in the suspended channel.

We demonstrated the capability to controllably tear the width of suspended graphene channels to tune their transport characteristics from the ballistic transport regime to the QD regime. In order to fully explore the structure and physics of suspended QDs made with this procedure, we will need to make even narrower devices, for instance by using narrower gold breakjunctions ($\sim$ 50 nm wide). The controlled cutting of suspended graphene channels into QDs, and possibly with ordered edges, offers exciting prospects for the development of spin qubits \cite{Engels13} and coherent electro-mechanical systems in graphene \cite{Weber14}.

In summary, we developed the capability to tailor 10-nm-scale suspended graphene junctions and quantum dot transistors. Using a low-power EM ($P_{Break} <$ 5 mW), we demonstrated the capability to make symmetric and low-disorder SGDs, whose channel length varies between 6 nm and 55 nm. We controlled their length by adjusting the power used during the gold EM. Furthermore, we could engineer the shape of the gold contacts to create highly-asymmetric source and drain electrodes by using a higher gold-EM power ($P_{Break} >$ 10 mW). Needled-shaped source electrodes and a non-uniform downward electrostatic force on the suspended graphene, allowed us to tear the channels along their length into very narrow junctions (down to 27 nm). Using low-temperature electron transport in these narrow junctions, we showed that their charge transport can be tuned from the ballistic to the QD regime. These ultra-short SGDs constitute a powerful new platform to explore the physics and applications of ballistic transport \cite{Fogler08}, strain-engineering \cite{Low10}, electro-mechanical coupling \cite{Waissman13, Weber14}, and ultra-high frequency NEMS \cite{Bunch07,Island12,Chen13}. This work was supported by NSERC (Canada), CFI (Canada), and Concordia University. We acknowledge usage of the QNI (Quebec Nano Infrastructure) cleanroom network.


\begin{thebibliography}{47}
\expandafter\ifx\csname natexlab\endcsname\relax\def\natexlab#1{#1}\fi
\expandafter\ifx\csname bibnamefont\endcsname\relax
  \def\bibnamefont#1{#1}\fi
\expandafter\ifx\csname bibfnamefont\endcsname\relax
  \def\bibfnamefont#1{#1}\fi
\expandafter\ifx\csname citenamefont\endcsname\relax
  \def\citenamefont#1{#1}\fi
\expandafter\ifx\csname url\endcsname\relax
  \def\url#1{\texttt{#1}}\fi
\expandafter\ifx\csname urlprefix\endcsname\relax\def\urlprefix{URL }\fi
\providecommand{\bibinfo}[2]{#2}
\providecommand{\eprint}[2][]{\url{#2}}

\bibitem[{\citenamefont{Meyer et~al.}(2008)\citenamefont{Meyer, Kisielowski,
  Erni, Rossell, Crommie, and Zettl}}]{Meyer08}
\bibinfo{author}{\bibfnamefont{J.~C.} \bibnamefont{Meyer}},
  \bibinfo{author}{\bibfnamefont{C.}~\bibnamefont{Kisielowski}},
  \bibinfo{author}{\bibfnamefont{R.}~\bibnamefont{Erni}},
  \bibinfo{author}{\bibfnamefont{M.~D.} \bibnamefont{Rossell}},
  \bibinfo{author}{\bibfnamefont{M.~F.} \bibnamefont{Crommie}},
  \bibnamefont{and} \bibinfo{author}{\bibfnamefont{A.}~\bibnamefont{Zettl}},
  \bibinfo{journal}{Nano Lett.} \textbf{\bibinfo{volume}{8}},
  \bibinfo{pages}{3582} (\bibinfo{year}{2008}).

\bibitem[{\citenamefont{Bolotin et~al.}(2008)\citenamefont{Bolotin, Sikes,
  Jiang, Klima, Fudenberg, Hone, Kim, and Stormer}}]{Bolotin08}
\bibinfo{author}{\bibfnamefont{K.~I.} \bibnamefont{Bolotin}},
  \bibinfo{author}{\bibfnamefont{K.~J.} \bibnamefont{Sikes}},
  \bibinfo{author}{\bibfnamefont{Z.}~\bibnamefont{Jiang}},
  \bibinfo{author}{\bibfnamefont{M.}~\bibnamefont{Klima}},
  \bibinfo{author}{\bibfnamefont{G.}~\bibnamefont{Fudenberg}},
  \bibinfo{author}{\bibfnamefont{J.}~\bibnamefont{Hone}},
  \bibinfo{author}{\bibfnamefont{P.}~\bibnamefont{Kim}}, \bibnamefont{and}
  \bibinfo{author}{\bibfnamefont{H.~L.} \bibnamefont{Stormer}},
  \bibinfo{journal}{Solid State Commun.} \textbf{\bibinfo{volume}{146}},
  \bibinfo{pages}{351} (\bibinfo{year}{2008}).

\bibitem[{\citenamefont{Du et~al.}(2008)\citenamefont{Du, Skachko, Barker, and
  Andrei}}]{Du08}
\bibinfo{author}{\bibfnamefont{X.}~\bibnamefont{Du}},
  \bibinfo{author}{\bibfnamefont{I.}~\bibnamefont{Skachko}},
  \bibinfo{author}{\bibfnamefont{A.}~\bibnamefont{Barker}}, \bibnamefont{and}
  \bibinfo{author}{\bibfnamefont{E.~Y.} \bibnamefont{Andrei}},
  \bibinfo{journal}{Nature Nanotechnol.} \textbf{\bibinfo{volume}{3}},
  \bibinfo{pages}{491} (\bibinfo{year}{2008}).

\bibitem[{\citenamefont{Yigen et~al.}(2013)\citenamefont{Yigen, Tayari, Island,
  Porter, and Champagne}}]{Yigen13}
\bibinfo{author}{\bibfnamefont{S.}~\bibnamefont{Yigen}},
  \bibinfo{author}{\bibfnamefont{V.}~\bibnamefont{Tayari}},
  \bibinfo{author}{\bibfnamefont{J.~O.} \bibnamefont{Island}},
  \bibinfo{author}{\bibfnamefont{J.~M.} \bibnamefont{Porter}},
  \bibnamefont{and} \bibinfo{author}{\bibfnamefont{A.~R.}
  \bibnamefont{Champagne}}, \bibinfo{journal}{Phys. Rev. B}
  \textbf{\bibinfo{volume}{87}}, \bibinfo{pages}{241411}
  (\bibinfo{year}{2013}).

\bibitem[{\citenamefont{Ponomarenko et~al.}(2008)\citenamefont{Ponomarenko,
  Schedin, Katsnelson, Yang, Hill, Novoselov, and Geim}}]{Ponomarenko08}
\bibinfo{author}{\bibfnamefont{L.~A.} \bibnamefont{Ponomarenko}},
  \bibinfo{author}{\bibfnamefont{F.}~\bibnamefont{Schedin}},
  \bibinfo{author}{\bibfnamefont{M.~I.} \bibnamefont{Katsnelson}},
  \bibinfo{author}{\bibfnamefont{R.}~\bibnamefont{Yang}},
  \bibinfo{author}{\bibfnamefont{E.~W.} \bibnamefont{Hill}},
  \bibinfo{author}{\bibfnamefont{K.~S.} \bibnamefont{Novoselov}},
  \bibnamefont{and} \bibinfo{author}{\bibfnamefont{A.~K.} \bibnamefont{Geim}},
  \bibinfo{journal}{Science} \textbf{\bibinfo{volume}{320}},
  \bibinfo{pages}{356} (\bibinfo{year}{2008}).

\bibitem[{\citenamefont{Young and Kim}(2009)}]{Young09}
\bibinfo{author}{\bibfnamefont{A.~F.} \bibnamefont{Young}} \bibnamefont{and}
  \bibinfo{author}{\bibfnamefont{P.}~\bibnamefont{Kim}},
  \bibinfo{journal}{Nature Phys.} \textbf{\bibinfo{volume}{5}},
  \bibinfo{pages}{222} (\bibinfo{year}{2009}).

\bibitem[{\citenamefont{Wu et~al.}(2012{\natexlab{a}})\citenamefont{Wu,
  Jenkins, Valdes-Garcia, Farmer, Zhu, Bol, Dimitrakopoulos, Zhu, Xia, Avouris
  et~al.}}]{Wu12}
\bibinfo{author}{\bibfnamefont{Y.~Q.} \bibnamefont{Wu}},
  \bibinfo{author}{\bibfnamefont{K.~A.} \bibnamefont{Jenkins}},
  \bibinfo{author}{\bibfnamefont{A.}~\bibnamefont{Valdes-Garcia}},
  \bibinfo{author}{\bibfnamefont{D.~B.} \bibnamefont{Farmer}},
  \bibinfo{author}{\bibfnamefont{Y.}~\bibnamefont{Zhu}},
  \bibinfo{author}{\bibfnamefont{A.~A.} \bibnamefont{Bol}},
  \bibinfo{author}{\bibfnamefont{C.}~\bibnamefont{Dimitrakopoulos}},
  \bibinfo{author}{\bibfnamefont{W.~J.} \bibnamefont{Zhu}},
  \bibinfo{author}{\bibfnamefont{F.~N.} \bibnamefont{Xia}},
  \bibinfo{author}{\bibfnamefont{P.}~\bibnamefont{Avouris}},
  \bibnamefont{et~al.}, \bibinfo{journal}{Nano Lett.}
  \textbf{\bibinfo{volume}{12}}, \bibinfo{pages}{3062}
  (\bibinfo{year}{2012}{\natexlab{a}}).

\bibitem[{\citenamefont{Guttinger et~al.}(2012)\citenamefont{Guttinger,
  Molitor, Stampfer, Schnez, Jacobsen, Droscher, Ihn, and
  Ensslin}}]{Guttinger12}
\bibinfo{author}{\bibfnamefont{J.}~\bibnamefont{Guttinger}},
  \bibinfo{author}{\bibfnamefont{F.}~\bibnamefont{Molitor}},
  \bibinfo{author}{\bibfnamefont{C.}~\bibnamefont{Stampfer}},
  \bibinfo{author}{\bibfnamefont{S.}~\bibnamefont{Schnez}},
  \bibinfo{author}{\bibfnamefont{A.}~\bibnamefont{Jacobsen}},
  \bibinfo{author}{\bibfnamefont{S.}~\bibnamefont{Droscher}},
  \bibinfo{author}{\bibfnamefont{T.}~\bibnamefont{Ihn}}, \bibnamefont{and}
  \bibinfo{author}{\bibfnamefont{K.}~\bibnamefont{Ensslin}},
  \bibinfo{journal}{Rep. Prog. Phys.} \textbf{\bibinfo{volume}{75}},
  \bibinfo{pages}{126502} (\bibinfo{year}{2012}).

\bibitem[{\citenamefont{Engels et~al.}(2013)\citenamefont{Engels, Epping, Volk,
  Korte, Voigtlander, Watanabe, Taniguchi, Trellenkamp, and
  Stampfer}}]{Engels13}
\bibinfo{author}{\bibfnamefont{S.}~\bibnamefont{Engels}},
  \bibinfo{author}{\bibfnamefont{A.}~\bibnamefont{Epping}},
  \bibinfo{author}{\bibfnamefont{C.}~\bibnamefont{Volk}},
  \bibinfo{author}{\bibfnamefont{S.}~\bibnamefont{Korte}},
  \bibinfo{author}{\bibfnamefont{B.}~\bibnamefont{Voigtlander}},
  \bibinfo{author}{\bibfnamefont{K.}~\bibnamefont{Watanabe}},
  \bibinfo{author}{\bibfnamefont{T.}~\bibnamefont{Taniguchi}},
  \bibinfo{author}{\bibfnamefont{S.}~\bibnamefont{Trellenkamp}},
  \bibnamefont{and} \bibinfo{author}{\bibfnamefont{C.}~\bibnamefont{Stampfer}},
  \bibinfo{journal}{Appl. Phys. Lett.} \textbf{\bibinfo{volume}{103}},
  \bibinfo{pages}{073113} (\bibinfo{year}{2013}).

\bibitem[{\citenamefont{Barreiro et~al.}(2012)\citenamefont{Barreiro, van~der
  Zant, and Vandersypen}}]{Barreiro12}
\bibinfo{author}{\bibfnamefont{A.}~\bibnamefont{Barreiro}},
  \bibinfo{author}{\bibfnamefont{H.~S.~J.} \bibnamefont{van~der Zant}},
  \bibnamefont{and} \bibinfo{author}{\bibfnamefont{L.~M.~K.}
  \bibnamefont{Vandersypen}}, \bibinfo{journal}{Nano Lett.}
  \textbf{\bibinfo{volume}{12}}, \bibinfo{pages}{6096} (\bibinfo{year}{2012}).

\bibitem[{\citenamefont{von Oppen et~al.}(2009)\citenamefont{von Oppen, Guinea,
  and Mariani}}]{vonOppen09}
\bibinfo{author}{\bibfnamefont{F.}~\bibnamefont{von Oppen}},
  \bibinfo{author}{\bibfnamefont{F.}~\bibnamefont{Guinea}}, \bibnamefont{and}
  \bibinfo{author}{\bibfnamefont{E.}~\bibnamefont{Mariani}},
  \bibinfo{journal}{Phys. Rev. B} \textbf{\bibinfo{volume}{80}},
  \bibinfo{pages}{075420} (\bibinfo{year}{2009}).

\bibitem[{\citenamefont{Weber et~al.}(2014)\citenamefont{Weber, Guttinger,
  Tsioutsios, Chang, and Bachtold}}]{Weber14}
\bibinfo{author}{\bibfnamefont{P.}~\bibnamefont{Weber}},
  \bibinfo{author}{\bibfnamefont{J.}~\bibnamefont{Guttinger}},
  \bibinfo{author}{\bibfnamefont{I.}~\bibnamefont{Tsioutsios}},
  \bibinfo{author}{\bibfnamefont{D.~E.} \bibnamefont{Chang}}, \bibnamefont{and}
  \bibinfo{author}{\bibfnamefont{A.}~\bibnamefont{Bachtold}},
  \bibinfo{journal}{Nano Lett.} \textbf{\bibinfo{volume}{14}},
  \bibinfo{pages}{2854} (\bibinfo{year}{2014}).

\bibitem[{\citenamefont{Shi et~al.}(2011)\citenamefont{Shi, Xu, Ralph, and
  McEuen}}]{Shi11}
\bibinfo{author}{\bibfnamefont{S.~F.} \bibnamefont{Shi}},
  \bibinfo{author}{\bibfnamefont{X.~D.} \bibnamefont{Xu}},
  \bibinfo{author}{\bibfnamefont{D.~C.} \bibnamefont{Ralph}}, \bibnamefont{and}
  \bibinfo{author}{\bibfnamefont{P.~L.} \bibnamefont{McEuen}},
  \bibinfo{journal}{Nano Lett.} \textbf{\bibinfo{volume}{11}},
  \bibinfo{pages}{1814} (\bibinfo{year}{2011}).

\bibitem[{\citenamefont{Fogler et~al.}(2008)\citenamefont{Fogler, Guinea, and
  Katsnelson}}]{Fogler08}
\bibinfo{author}{\bibfnamefont{M.~M.} \bibnamefont{Fogler}},
  \bibinfo{author}{\bibfnamefont{F.}~\bibnamefont{Guinea}}, \bibnamefont{and}
  \bibinfo{author}{\bibfnamefont{M.~I.} \bibnamefont{Katsnelson}},
  \bibinfo{journal}{Phys. Rev. Lett.} \textbf{\bibinfo{volume}{101}},
  \bibinfo{pages}{226804} (\bibinfo{year}{2008}).

\bibitem[{\citenamefont{Rocheleau et~al.}(2010)\citenamefont{Rocheleau, Ndukum,
  Macklin, Hertzberg, Clerk, and Schwab}}]{Rocheleau10}
\bibinfo{author}{\bibfnamefont{T.}~\bibnamefont{Rocheleau}},
  \bibinfo{author}{\bibfnamefont{T.}~\bibnamefont{Ndukum}},
  \bibinfo{author}{\bibfnamefont{C.}~\bibnamefont{Macklin}},
  \bibinfo{author}{\bibfnamefont{J.~B.} \bibnamefont{Hertzberg}},
  \bibinfo{author}{\bibfnamefont{A.~A.} \bibnamefont{Clerk}}, \bibnamefont{and}
  \bibinfo{author}{\bibfnamefont{K.~C.} \bibnamefont{Schwab}},
  \bibinfo{journal}{Nature} \textbf{\bibinfo{volume}{463}}, \bibinfo{pages}{72}
  (\bibinfo{year}{2010}).

\bibitem[{\citenamefont{Freitag et~al.}(2013)\citenamefont{Freitag, Low, and
  Avouris}}]{Freitag13}
\bibinfo{author}{\bibfnamefont{M.}~\bibnamefont{Freitag}},
  \bibinfo{author}{\bibfnamefont{T.}~\bibnamefont{Low}}, \bibnamefont{and}
  \bibinfo{author}{\bibfnamefont{P.}~\bibnamefont{Avouris}},
  \bibinfo{journal}{Nano Lett.} \textbf{\bibinfo{volume}{13}},
  \bibinfo{pages}{1644} (\bibinfo{year}{2013}).

\bibitem[{\citenamefont{Britnell et~al.}(2013)\citenamefont{Britnell, Ribeiro,
  Eckmann, Jalil, Belle, Mishchenko, Kim, Gorbachev, Georgiou, Morozov
  et~al.}}]{Britnell13}
\bibinfo{author}{\bibfnamefont{L.}~\bibnamefont{Britnell}},
  \bibinfo{author}{\bibfnamefont{R.~M.} \bibnamefont{Ribeiro}},
  \bibinfo{author}{\bibfnamefont{A.}~\bibnamefont{Eckmann}},
  \bibinfo{author}{\bibfnamefont{R.}~\bibnamefont{Jalil}},
  \bibinfo{author}{\bibfnamefont{B.~D.} \bibnamefont{Belle}},
  \bibinfo{author}{\bibfnamefont{A.}~\bibnamefont{Mishchenko}},
  \bibinfo{author}{\bibfnamefont{Y.~J.} \bibnamefont{Kim}},
  \bibinfo{author}{\bibfnamefont{R.~V.} \bibnamefont{Gorbachev}},
  \bibinfo{author}{\bibfnamefont{T.}~\bibnamefont{Georgiou}},
  \bibinfo{author}{\bibfnamefont{S.~V.} \bibnamefont{Morozov}},
  \bibnamefont{et~al.}, \bibinfo{journal}{Science}
  \textbf{\bibinfo{volume}{340}}, \bibinfo{pages}{1311} (\bibinfo{year}{2013}).

\bibitem[{\citenamefont{Kim et~al.}(2014)\citenamefont{Kim, Hwang, Kim, Shin,
  Kang, Kim, Jang, Kim, Lee, Choi et~al.}}]{Kim14}
\bibinfo{author}{\bibfnamefont{C.~O.} \bibnamefont{Kim}},
  \bibinfo{author}{\bibfnamefont{S.~W.} \bibnamefont{Hwang}},
  \bibinfo{author}{\bibfnamefont{S.}~\bibnamefont{Kim}},
  \bibinfo{author}{\bibfnamefont{D.~H.} \bibnamefont{Shin}},
  \bibinfo{author}{\bibfnamefont{S.~S.} \bibnamefont{Kang}},
  \bibinfo{author}{\bibfnamefont{J.~M.} \bibnamefont{Kim}},
  \bibinfo{author}{\bibfnamefont{C.~W.} \bibnamefont{Jang}},
  \bibinfo{author}{\bibfnamefont{J.~H.} \bibnamefont{Kim}},
  \bibinfo{author}{\bibfnamefont{K.~W.} \bibnamefont{Lee}},
  \bibinfo{author}{\bibfnamefont{S.~H.} \bibnamefont{Choi}},
  \bibnamefont{et~al.}, \bibinfo{journal}{Scien. Rep.}
  \textbf{\bibinfo{volume}{4}}, \bibinfo{pages}{5603} (\bibinfo{year}{2014}).

\bibitem[{\citenamefont{Guinea et~al.}(2010)\citenamefont{Guinea, Katsnelson,
  and Geim}}]{Guinea10}
\bibinfo{author}{\bibfnamefont{F.}~\bibnamefont{Guinea}},
  \bibinfo{author}{\bibfnamefont{M.~I.} \bibnamefont{Katsnelson}},
  \bibnamefont{and} \bibinfo{author}{\bibfnamefont{A.~K.} \bibnamefont{Geim}},
  \bibinfo{journal}{Nature Phys.} \textbf{\bibinfo{volume}{6}},
  \bibinfo{pages}{30} (\bibinfo{year}{2010}).

\bibitem[{\citenamefont{Low and Guinea}(2010)}]{Low10}
\bibinfo{author}{\bibfnamefont{T.}~\bibnamefont{Low}} \bibnamefont{and}
  \bibinfo{author}{\bibfnamefont{F.}~\bibnamefont{Guinea}},
  \bibinfo{journal}{Nano Lett.} \textbf{\bibinfo{volume}{10}},
  \bibinfo{pages}{3551} (\bibinfo{year}{2010}).

\bibitem[{\citenamefont{Champagne et~al.}(2005)\citenamefont{Champagne,
  Pasupathy, and Ralph}}]{Champagne05}
\bibinfo{author}{\bibfnamefont{A.~R.} \bibnamefont{Champagne}},
  \bibinfo{author}{\bibfnamefont{A.~N.} \bibnamefont{Pasupathy}},
  \bibnamefont{and} \bibinfo{author}{\bibfnamefont{D.~C.} \bibnamefont{Ralph}},
  \bibinfo{journal}{Nano Lett.} \textbf{\bibinfo{volume}{5}},
  \bibinfo{pages}{305} (\bibinfo{year}{2005}).

\bibitem[{\citenamefont{Parks et~al.}(2010)\citenamefont{Parks, Champagne,
  Costi, Shum, Pasupathy, Neuscamman, Flores-Torres, Cornaglia, Aligia,
  Balseiro et~al.}}]{Parks10}
\bibinfo{author}{\bibfnamefont{J.~J.} \bibnamefont{Parks}},
  \bibinfo{author}{\bibfnamefont{A.~R.} \bibnamefont{Champagne}},
  \bibinfo{author}{\bibfnamefont{T.~A.} \bibnamefont{Costi}},
  \bibinfo{author}{\bibfnamefont{W.~W.} \bibnamefont{Shum}},
  \bibinfo{author}{\bibfnamefont{A.~N.} \bibnamefont{Pasupathy}},
  \bibinfo{author}{\bibfnamefont{E.}~\bibnamefont{Neuscamman}},
  \bibinfo{author}{\bibfnamefont{S.}~\bibnamefont{Flores-Torres}},
  \bibinfo{author}{\bibfnamefont{P.~S.} \bibnamefont{Cornaglia}},
  \bibinfo{author}{\bibfnamefont{A.~A.} \bibnamefont{Aligia}},
  \bibinfo{author}{\bibfnamefont{C.~A.} \bibnamefont{Balseiro}},
  \bibnamefont{et~al.}, \bibinfo{journal}{Science}
  \textbf{\bibinfo{volume}{328}}, \bibinfo{pages}{1370} (\bibinfo{year}{2010}).

\bibitem[{\citenamefont{Zheng et~al.}(2013)\citenamefont{Zheng, Wang, Quhe,
  Liu, Li, Yu, Mei, Shi, Gao, and Lu}}]{Zheng13}
\bibinfo{author}{\bibfnamefont{J.~X.} \bibnamefont{Zheng}},
  \bibinfo{author}{\bibfnamefont{L.}~\bibnamefont{Wang}},
  \bibinfo{author}{\bibfnamefont{R.~G.} \bibnamefont{Quhe}},
  \bibinfo{author}{\bibfnamefont{Q.~H.} \bibnamefont{Liu}},
  \bibinfo{author}{\bibfnamefont{H.}~\bibnamefont{Li}},
  \bibinfo{author}{\bibfnamefont{D.~P.} \bibnamefont{Yu}},
  \bibinfo{author}{\bibfnamefont{W.~N.} \bibnamefont{Mei}},
  \bibinfo{author}{\bibfnamefont{J.~J.} \bibnamefont{Shi}},
  \bibinfo{author}{\bibfnamefont{Z.~X.} \bibnamefont{Gao}}, \bibnamefont{and}
  \bibinfo{author}{\bibfnamefont{J.}~\bibnamefont{Lu}},
  \bibinfo{journal}{Scien. Rep.} \textbf{\bibinfo{volume}{3}},
  \bibinfo{pages}{1314} (\bibinfo{year}{2013}).

\bibitem[{\citenamefont{Fong et~al.}(2013)\citenamefont{Fong, Wollman, Ravi,
  Chen, Clerk, Shaw, Leduc, and Schwab}}]{Fong13}
\bibinfo{author}{\bibfnamefont{K.~C.} \bibnamefont{Fong}},
  \bibinfo{author}{\bibfnamefont{E.~E.} \bibnamefont{Wollman}},
  \bibinfo{author}{\bibfnamefont{H.}~\bibnamefont{Ravi}},
  \bibinfo{author}{\bibfnamefont{W.}~\bibnamefont{Chen}},
  \bibinfo{author}{\bibfnamefont{A.~A.} \bibnamefont{Clerk}},
  \bibinfo{author}{\bibfnamefont{M.~D.} \bibnamefont{Shaw}},
  \bibinfo{author}{\bibfnamefont{H.~G.} \bibnamefont{Leduc}}, \bibnamefont{and}
  \bibinfo{author}{\bibfnamefont{K.~C.} \bibnamefont{Schwab}},
  \bibinfo{journal}{Physical Review X} \textbf{\bibinfo{volume}{3}}
  (\bibinfo{year}{2013}).

\bibitem[{\citenamefont{Kim et~al.}(2011)\citenamefont{Kim, Choi, Kim, Cho, and
  Chung}}]{Kim11}
\bibinfo{author}{\bibfnamefont{K.}~\bibnamefont{Kim}},
  \bibinfo{author}{\bibfnamefont{J.~Y.} \bibnamefont{Choi}},
  \bibinfo{author}{\bibfnamefont{T.}~\bibnamefont{Kim}},
  \bibinfo{author}{\bibfnamefont{S.~H.} \bibnamefont{Cho}}, \bibnamefont{and}
  \bibinfo{author}{\bibfnamefont{H.~J.} \bibnamefont{Chung}},
  \bibinfo{journal}{Nature} \textbf{\bibinfo{volume}{479}},
  \bibinfo{pages}{338} (\bibinfo{year}{2011}).

\bibitem[{\citenamefont{Wu et~al.}(2012{\natexlab{b}})\citenamefont{Wu,
  Perebeinos, Lin, Low, Xia, and Avouris}}]{Wu12_2}
\bibinfo{author}{\bibfnamefont{Y.~Q.} \bibnamefont{Wu}},
  \bibinfo{author}{\bibfnamefont{V.}~\bibnamefont{Perebeinos}},
  \bibinfo{author}{\bibfnamefont{Y.~M.} \bibnamefont{Lin}},
  \bibinfo{author}{\bibfnamefont{T.}~\bibnamefont{Low}},
  \bibinfo{author}{\bibfnamefont{F.~N.} \bibnamefont{Xia}}, \bibnamefont{and}
  \bibinfo{author}{\bibfnamefont{P.}~\bibnamefont{Avouris}},
  \bibinfo{journal}{Nano Lett.} \textbf{\bibinfo{volume}{12}},
  \bibinfo{pages}{1417} (\bibinfo{year}{2012}{\natexlab{b}}).

\bibitem[{\citenamefont{Moser and Bachtold}(2009)}]{Moser09}
\bibinfo{author}{\bibfnamefont{J.}~\bibnamefont{Moser}} \bibnamefont{and}
  \bibinfo{author}{\bibfnamefont{A.}~\bibnamefont{Bachtold}},
  \bibinfo{journal}{Appl. Phys. Lett.} \textbf{\bibinfo{volume}{95}},
  \bibinfo{pages}{173506} (\bibinfo{year}{2009}).

\bibitem[{\citenamefont{Lu et~al.}(2010)\citenamefont{Lu, Goldsmith, Strachan,
  Lim, Luo, and Johnson}}]{Lu10}
\bibinfo{author}{\bibfnamefont{Y.}~\bibnamefont{Lu}},
  \bibinfo{author}{\bibfnamefont{B.}~\bibnamefont{Goldsmith}},
  \bibinfo{author}{\bibfnamefont{D.~R.} \bibnamefont{Strachan}},
  \bibinfo{author}{\bibfnamefont{J.~H.} \bibnamefont{Lim}},
  \bibinfo{author}{\bibfnamefont{Z.~T.} \bibnamefont{Luo}}, \bibnamefont{and}
  \bibinfo{author}{\bibfnamefont{A.~T.~C.} \bibnamefont{Johnson}},
  \bibinfo{journal}{Small} \textbf{\bibinfo{volume}{6}}, \bibinfo{pages}{2748}
  (\bibinfo{year}{2010}).

\bibitem[{\citenamefont{Giovannetti et~al.}(2008)\citenamefont{Giovannetti,
  Khomyakov, Brocks, Karpan, van~den Brink, and Kelly}}]{Giovannetti08}
\bibinfo{author}{\bibfnamefont{G.}~\bibnamefont{Giovannetti}},
  \bibinfo{author}{\bibfnamefont{P.~A.} \bibnamefont{Khomyakov}},
  \bibinfo{author}{\bibfnamefont{G.}~\bibnamefont{Brocks}},
  \bibinfo{author}{\bibfnamefont{V.~M.} \bibnamefont{Karpan}},
  \bibinfo{author}{\bibfnamefont{J.}~\bibnamefont{van~den Brink}},
  \bibnamefont{and} \bibinfo{author}{\bibfnamefont{P.~J.} \bibnamefont{Kelly}},
  \bibinfo{journal}{Phys. Rev. Lett.} \textbf{\bibinfo{volume}{101}},
  \bibinfo{pages}{026803} (\bibinfo{year}{2008}).

\bibitem[{\citenamefont{Golizadeh-Mojarad and Datta}(2009)}]{Golizadeh09}
\bibinfo{author}{\bibfnamefont{R.}~\bibnamefont{Golizadeh-Mojarad}}
  \bibnamefont{and} \bibinfo{author}{\bibfnamefont{S.}~\bibnamefont{Datta}},
  \bibinfo{journal}{Phys. Rev. B} \textbf{\bibinfo{volume}{79}},
  \bibinfo{pages}{085410} (\bibinfo{year}{2009}).

\bibitem[{\citenamefont{Tworzydlo et~al.}(2006)\citenamefont{Tworzydlo,
  Trauzettel, Titov, Rycerz, and Beenakker}}]{Tworzydlo06}
\bibinfo{author}{\bibfnamefont{J.}~\bibnamefont{Tworzydlo}},
  \bibinfo{author}{\bibfnamefont{B.}~\bibnamefont{Trauzettel}},
  \bibinfo{author}{\bibfnamefont{M.}~\bibnamefont{Titov}},
  \bibinfo{author}{\bibfnamefont{A.}~\bibnamefont{Rycerz}}, \bibnamefont{and}
  \bibinfo{author}{\bibfnamefont{C.~W.~J.} \bibnamefont{Beenakker}},
  \bibinfo{journal}{Phys. Rev. Lett.} \textbf{\bibinfo{volume}{96}},
  \bibinfo{pages}{246802} (\bibinfo{year}{2006}).

\bibitem[{\citenamefont{Han et~al.}(2007)\citenamefont{Han, Ozyilmaz, Zhang,
  and Kim}}]{Han07}
\bibinfo{author}{\bibfnamefont{M.~Y.} \bibnamefont{Han}},
  \bibinfo{author}{\bibfnamefont{B.}~\bibnamefont{Ozyilmaz}},
  \bibinfo{author}{\bibfnamefont{Y.~B.} \bibnamefont{Zhang}}, \bibnamefont{and}
  \bibinfo{author}{\bibfnamefont{P.}~\bibnamefont{Kim}},
  \bibinfo{journal}{Phys. Rev. Lett.} \textbf{\bibinfo{volume}{98}},
  \bibinfo{pages}{206805} (\bibinfo{year}{2007}).

\bibitem[{\citenamefont{Moura and Marder}(2013)}]{Moura13}
\bibinfo{author}{\bibfnamefont{M.~J.~B.} \bibnamefont{Moura}} \bibnamefont{and}
  \bibinfo{author}{\bibfnamefont{M.}~\bibnamefont{Marder}},
  \bibinfo{journal}{Phys. Rev. E} \textbf{\bibinfo{volume}{88}},
  \bibinfo{pages}{032405} (\bibinfo{year}{2013}).

\bibitem[{\citenamefont{Park et~al.}(1999)\citenamefont{Park, Lim, Alivisatos,
  Park, and McEuen}}]{Park99}
\bibinfo{author}{\bibfnamefont{H.}~\bibnamefont{Park}},
  \bibinfo{author}{\bibfnamefont{A.~K.~L.} \bibnamefont{Lim}},
  \bibinfo{author}{\bibfnamefont{A.~P.} \bibnamefont{Alivisatos}},
  \bibinfo{author}{\bibfnamefont{J.}~\bibnamefont{Park}}, \bibnamefont{and}
  \bibinfo{author}{\bibfnamefont{P.~L.} \bibnamefont{McEuen}},
  \bibinfo{journal}{Appl. Phys. Lett.} \textbf{\bibinfo{volume}{75}},
  \bibinfo{pages}{301} (\bibinfo{year}{1999}).

\bibitem[{\citenamefont{Esen and Fuhrer}(2005)}]{Esen05}
\bibinfo{author}{\bibfnamefont{G.}~\bibnamefont{Esen}} \bibnamefont{and}
  \bibinfo{author}{\bibfnamefont{M.~S.} \bibnamefont{Fuhrer}},
  \bibinfo{journal}{Appl. Phys. Lett.} \textbf{\bibinfo{volume}{87}},
  \bibinfo{pages}{263101} (\bibinfo{year}{2005}).

\bibitem[{\citenamefont{Island et~al.}(2011)\citenamefont{Island, Tayari,
  Yigen, McRae, and Champagne}}]{Island11}
\bibinfo{author}{\bibfnamefont{J.~O.} \bibnamefont{Island}},
  \bibinfo{author}{\bibfnamefont{V.}~\bibnamefont{Tayari}},
  \bibinfo{author}{\bibfnamefont{S.}~\bibnamefont{Yigen}},
  \bibinfo{author}{\bibfnamefont{A.~C.} \bibnamefont{McRae}}, \bibnamefont{and}
  \bibinfo{author}{\bibfnamefont{A.~R.} \bibnamefont{Champagne}},
  \bibinfo{journal}{Appl. Phys. Lett.} \textbf{\bibinfo{volume}{99}},
  \bibinfo{pages}{243106} (\bibinfo{year}{2011}).

\bibitem[{\citenamefont{Island et~al.}(2012)\citenamefont{Island, Tayari,
  McRae, and Champagne}}]{Island12}
\bibinfo{author}{\bibfnamefont{J.~O.} \bibnamefont{Island}},
  \bibinfo{author}{\bibfnamefont{V.}~\bibnamefont{Tayari}},
  \bibinfo{author}{\bibfnamefont{A.~C.} \bibnamefont{McRae}}, \bibnamefont{and}
  \bibinfo{author}{\bibfnamefont{A.~R.} \bibnamefont{Champagne}},
  \bibinfo{journal}{Nano Lett.} \textbf{\bibinfo{volume}{12}},
  \bibinfo{pages}{4564} (\bibinfo{year}{2012}).

\bibitem[{\citenamefont{Jeong et~al.}(2014)\citenamefont{Jeong, Kim, Kim, Lee,
  and Reddy}}]{Jeong14}
\bibinfo{author}{\bibfnamefont{W.}~\bibnamefont{Jeong}},
  \bibinfo{author}{\bibfnamefont{K.}~\bibnamefont{Kim}},
  \bibinfo{author}{\bibfnamefont{Y.}~\bibnamefont{Kim}},
  \bibinfo{author}{\bibfnamefont{W.}~\bibnamefont{Lee}}, \bibnamefont{and}
  \bibinfo{author}{\bibfnamefont{P.}~\bibnamefont{Reddy}},
  \bibinfo{journal}{Scien. Rep.} \textbf{\bibinfo{volume}{4}},
  \bibinfo{pages}{4975} (\bibinfo{year}{2014}).

\bibitem[{\citenamefont{Dorgan et~al.}(2013)\citenamefont{Dorgan, Behnam,
  Conley, Bolotin, and Pop}}]{Dorgan13}
\bibinfo{author}{\bibfnamefont{V.~E.} \bibnamefont{Dorgan}},
  \bibinfo{author}{\bibfnamefont{A.}~\bibnamefont{Behnam}},
  \bibinfo{author}{\bibfnamefont{H.~J.} \bibnamefont{Conley}},
  \bibinfo{author}{\bibfnamefont{K.~I.} \bibnamefont{Bolotin}},
  \bibnamefont{and} \bibinfo{author}{\bibfnamefont{E.}~\bibnamefont{Pop}},
  \bibinfo{journal}{Nano Lett.} \textbf{\bibinfo{volume}{13}},
  \bibinfo{pages}{4581} (\bibinfo{year}{2013}).

\bibitem[{\citenamefont{Das~Sarma et~al.}(2011)\citenamefont{Das~Sarma, Adam,
  Hwang, and Rossi}}]{DasSarma11}
\bibinfo{author}{\bibfnamefont{S.}~\bibnamefont{Das~Sarma}},
  \bibinfo{author}{\bibfnamefont{S.}~\bibnamefont{Adam}},
  \bibinfo{author}{\bibfnamefont{E.~H.} \bibnamefont{Hwang}}, \bibnamefont{and}
  \bibinfo{author}{\bibfnamefont{E.}~\bibnamefont{Rossi}},
  \bibinfo{journal}{Rev. Mod. Phys.} \textbf{\bibinfo{volume}{83}},
  \bibinfo{pages}{407} (\bibinfo{year}{2011}).

\bibitem[{\citenamefont{Tan et~al.}(2007)\citenamefont{Tan, Zhang, Bolotin,
  Zhao, Adam, Hwang, Das~Sarma, Stormer, and Kim}}]{Tan07}
\bibinfo{author}{\bibfnamefont{Y.~W.} \bibnamefont{Tan}},
  \bibinfo{author}{\bibfnamefont{Y.}~\bibnamefont{Zhang}},
  \bibinfo{author}{\bibfnamefont{K.}~\bibnamefont{Bolotin}},
  \bibinfo{author}{\bibfnamefont{Y.}~\bibnamefont{Zhao}},
  \bibinfo{author}{\bibfnamefont{S.}~\bibnamefont{Adam}},
  \bibinfo{author}{\bibfnamefont{E.~H.} \bibnamefont{Hwang}},
  \bibinfo{author}{\bibfnamefont{S.}~\bibnamefont{Das~Sarma}},
  \bibinfo{author}{\bibfnamefont{H.~L.} \bibnamefont{Stormer}},
  \bibnamefont{and} \bibinfo{author}{\bibfnamefont{P.}~\bibnamefont{Kim}},
  \bibinfo{journal}{Phys. Rev. Lett.} \textbf{\bibinfo{volume}{99}},
  \bibinfo{pages}{246803} (\bibinfo{year}{2007}).

\bibitem[{\citenamefont{Bunch et~al.}(2007)\citenamefont{Bunch, van~der Zande,
  Verbridge, Frank, Tanenbaum, Parpia, Craighead, and McEuen}}]{Bunch07}
\bibinfo{author}{\bibfnamefont{J.~S.} \bibnamefont{Bunch}},
  \bibinfo{author}{\bibfnamefont{A.~M.} \bibnamefont{van~der Zande}},
  \bibinfo{author}{\bibfnamefont{S.~S.} \bibnamefont{Verbridge}},
  \bibinfo{author}{\bibfnamefont{I.~W.} \bibnamefont{Frank}},
  \bibinfo{author}{\bibfnamefont{D.~M.} \bibnamefont{Tanenbaum}},
  \bibinfo{author}{\bibfnamefont{J.~M.} \bibnamefont{Parpia}},
  \bibinfo{author}{\bibfnamefont{H.~G.} \bibnamefont{Craighead}},
  \bibnamefont{and} \bibinfo{author}{\bibfnamefont{P.~L.}
  \bibnamefont{McEuen}}, \bibinfo{journal}{Science}
  \textbf{\bibinfo{volume}{315}}, \bibinfo{pages}{490} (\bibinfo{year}{2007}).

\bibitem[{\citenamefont{Chen et~al.}(2013)\citenamefont{Chen, Lee, Deshpande,
  Lee, Lekas, Shepard, and Hone}}]{Chen13}
\bibinfo{author}{\bibfnamefont{C.~Y.} \bibnamefont{Chen}},
  \bibinfo{author}{\bibfnamefont{S.}~\bibnamefont{Lee}},
  \bibinfo{author}{\bibfnamefont{V.~V.} \bibnamefont{Deshpande}},
  \bibinfo{author}{\bibfnamefont{G.~H.} \bibnamefont{Lee}},
  \bibinfo{author}{\bibfnamefont{M.}~\bibnamefont{Lekas}},
  \bibinfo{author}{\bibfnamefont{K.}~\bibnamefont{Shepard}}, \bibnamefont{and}
  \bibinfo{author}{\bibfnamefont{J.}~\bibnamefont{Hone}},
  \bibinfo{journal}{Nature Nanotechnol.} \textbf{\bibinfo{volume}{8}},
  \bibinfo{pages}{923} (\bibinfo{year}{2013}).

\bibitem[{\citenamefont{Goldhaber-Gordon
  et~al.}(1998)\citenamefont{Goldhaber-Gordon, Gores, Kastner, Shtrikman,
  Mahalu, and Meirav}}]{Goldhaber98}
\bibinfo{author}{\bibfnamefont{D.}~\bibnamefont{Goldhaber-Gordon}},
  \bibinfo{author}{\bibfnamefont{J.}~\bibnamefont{Gores}},
  \bibinfo{author}{\bibfnamefont{M.~A.} \bibnamefont{Kastner}},
  \bibinfo{author}{\bibfnamefont{H.}~\bibnamefont{Shtrikman}},
  \bibinfo{author}{\bibfnamefont{D.}~\bibnamefont{Mahalu}}, \bibnamefont{and}
  \bibinfo{author}{\bibfnamefont{U.}~\bibnamefont{Meirav}},
  \bibinfo{journal}{Phys. Rev. Lett.} \textbf{\bibinfo{volume}{81}},
  \bibinfo{pages}{5225} (\bibinfo{year}{1998}).

\bibitem[{\citenamefont{Wang et~al.}(2011)\citenamefont{Wang, Ouyang, Jiao,
  Wang, Xie, Wu, Guo, and Dai}}]{Wang11}
\bibinfo{author}{\bibfnamefont{X.~R.} \bibnamefont{Wang}},
  \bibinfo{author}{\bibfnamefont{Y.~J.} \bibnamefont{Ouyang}},
  \bibinfo{author}{\bibfnamefont{L.~Y.} \bibnamefont{Jiao}},
  \bibinfo{author}{\bibfnamefont{H.~L.} \bibnamefont{Wang}},
  \bibinfo{author}{\bibfnamefont{L.~M.} \bibnamefont{Xie}},
  \bibinfo{author}{\bibfnamefont{J.}~\bibnamefont{Wu}},
  \bibinfo{author}{\bibfnamefont{J.}~\bibnamefont{Guo}}, \bibnamefont{and}
  \bibinfo{author}{\bibfnamefont{H.~J.} \bibnamefont{Dai}},
  \bibinfo{journal}{Nature Nanotechnol.} \textbf{\bibinfo{volume}{6}},
  \bibinfo{pages}{563} (\bibinfo{year}{2011}).

\bibitem[{\citenamefont{Yang et~al.}(2007)\citenamefont{Yang, Park, Son, Cohen,
  and Louie}}]{Yang07}
\bibinfo{author}{\bibfnamefont{L.}~\bibnamefont{Yang}},
  \bibinfo{author}{\bibfnamefont{C.~H.} \bibnamefont{Park}},
  \bibinfo{author}{\bibfnamefont{Y.~W.} \bibnamefont{Son}},
  \bibinfo{author}{\bibfnamefont{M.~L.} \bibnamefont{Cohen}}, \bibnamefont{and}
  \bibinfo{author}{\bibfnamefont{S.~G.} \bibnamefont{Louie}},
  \bibinfo{journal}{Phys. Rev. Lett.} \textbf{\bibinfo{volume}{99}},
  \bibinfo{pages}{186801} (\bibinfo{year}{2007}).

\bibitem[{\citenamefont{Waissman et~al.}(2013)\citenamefont{Waissman, Honig,
  Pecker, Benyamini, Hamo, and Ilani}}]{Waissman13}
\bibinfo{author}{\bibfnamefont{J.}~\bibnamefont{Waissman}},
  \bibinfo{author}{\bibfnamefont{M.}~\bibnamefont{Honig}},
  \bibinfo{author}{\bibfnamefont{S.}~\bibnamefont{Pecker}},
  \bibinfo{author}{\bibfnamefont{A.}~\bibnamefont{Benyamini}},
  \bibinfo{author}{\bibfnamefont{A.}~\bibnamefont{Hamo}}, \bibnamefont{and}
  \bibinfo{author}{\bibfnamefont{S.}~\bibnamefont{Ilani}},
  \bibinfo{journal}{Nature Nanotechnol.} \textbf{\bibinfo{volume}{8}},
  \bibinfo{pages}{569} (\bibinfo{year}{2013}).

\end{thebibliography}
\end{document}